\documentclass[12pt,a4paper]{article}
\usepackage{amsfonts,latexsym}
\usepackage{amsmath,amssymb}
\usepackage{graphicx,color}

\renewcommand{\thefootnote}{\fnsymbol{footnote}}

\begin{document}

\vspace{12mm}

\begin{center}
{{{\Large {\bf Shadow and geometric scattering  analysis  of a new black hole with magnetic charge }}}}\\[10mm]

{Yun Soo Myung\footnote{e-mail address: ysmyung@inje.ac.kr}}\\[8mm]

{Center for Quantum Spacetime, Sogang University, Seoul 04107, Republic of  Korea\\[0pt] }

\end{center}
\vspace{2mm}

\begin{abstract}
We obtain a newly charged  black hole  with magnetic charge $q$ and coupling constant $\mu$ from the Einstein-nonlinear electrodynamics theory inspired by quasi-topological terms.
We perform the shadow and geometric scattering  analysis of this  black hole.
For $\mu=0.01$,  we have four solution branches labelled by $2\pm$, 3, and  4 of the horizon, while  there exists  a single branch labelled by 2 for $\mu=0.1$.
There is the naked singularity (NS) arisen  from the magnetic charge extension of the  photon spheres for the 3 and 4-branches.
In case of $q<1$, shadow radii of the $2$-branch with $\mu$=0.01 and 0.1 are the same as that of the Reissner-Nordstr\"om black hole,  while  the 3-NS branch of $q>1$  is constrained by the EHT observation.
The  geometric scattering  analysis is performed to understand the peculiar forms of the critical impact factors.
\end{abstract}
\vspace{5mm}

\vspace{1.5cm}

\hspace{11.5cm}{Typeset Using \LaTeX}
\newpage
\renewcommand{\thefootnote}{\arabic{footnote}}
\setcounter{footnote}{0}

%%%% Introduction %%%%

\section{Introduction}
Black holes are considered as the most mysterious object of celestial bodies obtained from general relativity.
It is believed that supermassive black holes founded at the center of galaxies have played the important role in galaxy formation and galaxy evolution.
 The images of the M87$^*$ black hole (BH)~\cite{EventHorizonTelescope:2019dse,EventHorizonTelescope:2019ths,EventHorizonTelescope:2019ggy} have inspired enormous studies on the  BH.
Recently, the EHT observation has focussed on  the center of our galaxy and delivered   promising  images of the  SgrA$^*$ BH~\cite{EventHorizonTelescope:2022wkp,EventHorizonTelescope:2022wok,EventHorizonTelescope:2022xqj}.

No-hair theorem states that the BH is completely described by the mass, charge, and rotation parameter~\cite{Ruffini:1971bza}, but the scalar hair would be an emergent hair~\cite{Herdeiro:2015waa}.
The shadow of BH with scalar hair was used  to test the EHT results~\cite{Khodadi:2020jij}, whereas the shadows of other BHs, wormholes, and naked singularities  obtained  from various modified gravity theories  have been employed  to constrain their hair parameters~\cite{Vagnozzi:2022moj}.

Other  BH  solutions with magnetic/elelctric charges, differing from Reissner-Nordstr\"om black hole (RNBH)  were found from the Einstein-Euler-Heisenberg-nonlinear electrodynamics (NED) theory~\cite{Yajima:2000kw,Amaro:2020xro,Breton:2021mju,Allahyari:2019jqz}.
Recently, these BHs were studied for thermodynamics~\cite{Magos:2020ykt,Zhao:2024phz,Gursel:2025wan}, weak deflection angle~\cite{Fu:2021akc}, optical appearance~\cite{Zeng:2022pvb}, and  thermodynamic and shadow radius analysis~\cite{Myung:2025zxu}. Also, the dyonic BH solution was obtained from Einstein-Maxwell theory with a quasi-topological term~\cite{Liu:2019rib,Cisterna:2020rkc} and its scalarized black holes were extensively investigated in~\cite{Myung:2020ctt}.

In this work, we wish to obtain a newly charged  black hole  with magnetic charge $q$ and coupling constant $\mu$ from the Einstein-NED theory inspired by quasi-topological terms.
We  will perform  shadow radius and geometric scattering analysis of this black hole.
Our analysis depends critically  on the coupling constant $\mu$.
 For $\mu=0.01<0.021$,  we have four solution branches of the horizon including $2\pm$, 3, and  4-branches.
 There is the naked singularity (NS) arisen  from the charge extension of the  photon spheres for the 3, 4-branches.
 It is interesting  to note that  for $\mu=0.1>0.021$, there is no limitation on the magnetic charge $q$ for its horizon, photon sphere, and shadow radius.
  Therefore, there is no NS arisen  from the charge extension of the  photon sphere for $\mu=0.1$.   The shadow radii for the 2-branch with $\mu$=0.01 and 0.1 are  the  same as that for the RNBH, while  the $q>1$ 3-branch   is constrained by the recent EHT observation. We would like to mention that  the analysis of shadow radius   was discussed by including the magnetically charged BH~\cite{Allahyari:2019jqz,Vagnozzi:2022moj}.
Finally,  we perform the  geometric scattering  analysis  to understand the peculiar forms of the critical impact factors $b_3(m=1,q,\mu=0.01)$ and $b_{2+}(1,q+,0.01)$.

\section{A newly charged BH  solution}
The quasi-topological term has no effect on the purely electric
or magnetic RNBH solution, but the dyonic solution is completely
modified.
As was shown in~\cite{Liu:2019rib}, a few of low lying  quasi-topological terms are given by
\begin{eqnarray}
&&U^{(1)}=-\mathcal{F}, \label{t0-terms} \\
&&U^{(2)}=-2F^{(4)}+\mathcal{F}^2, \label{t1-terms} \\
&&U^{(3)}=-8F^{(6)}-6\mathcal{F}F^{(4)}+\mathcal{F}^3 \label{t2-terms}
\end{eqnarray}
 with
 \begin{equation}
 \mathcal{F}=F^{\mu\nu}F_{\mu\nu}\quad F^{(4)}=F^\mu_\nu F^\nu_\rho F^\rho_\sigma F^\sigma_\mu,\quad  F^{(6)}=F^\mu_\nu F^\nu_\rho F^\rho_\sigma F^\sigma_\kappa F^\kappa_\eta F^\eta_\mu.
 \end{equation}
For a dyonic solution, $U^{(2)}$ is  non-vanishing but  it vanishes for each electrically (magnetically) charged RNBH solution.
In this study, we would like to choose  a single term $\mathcal{F}^3$  of $U^{(3)}$ to find a new magnetically charged BH solution. The other choice of $F^{(6)}$ or $\mathcal{F}F^{(4)}$  will provide the same result with a different coefficient in front of  $q^6/r^{10}$.  If one chooses $\mathcal{F}^2$ in $U^{(2)}$, it gave us a magnetically charged BH solution (\ref{g-func})~\cite{Yajima:2000kw}.

We introduce  the  Einstein-NED (ENED) theory inspired by quasi-topological terms
\begin{equation}
{\cal L}_{\rm ENED}=\frac{1}{16\pi}\Big[R-(\mathcal{F}-\mu\mathcal{F}^3)\Big],\label{EEH-a}
\end{equation}
with a coupling constant $\mu$.
Two equations are derived from the above action as
\begin{eqnarray}
&&G_{\mu\nu}=2\Big(F_{\mu\rho}F_\nu~^\rho-\frac{\mathcal{F}}{4}g_{\mu\nu}-3\mu\mathcal{F}^2F_{\mu\rho}F_\nu~^\rho+\frac{\mu \mathcal{F}^3}{4}g_{\mu\nu}\Big), \label{eq-1} \\
&&\nabla_\mu(F^{\mu\nu}-3\mu \mathcal{F}^2F^{\mu\nu})=0.\label{eq-2}
\end{eqnarray}

To find a newly magnetically charged  black hole (NMBH) solution, we consider  a spherically symmetric  spacetime with  the gauge potential $A_{\varphi}=-q \cos \theta$  as
\begin{equation}\label{metric-ansatz}
ds^2_{\rm NMBH}=-f(r)dt^2+\frac{dr^2}{f(r)}+r^2(d\theta^2+\sin^2\theta d\phi^2),\quad \mathcal{F}=\frac{2q^2}{r^4}.
\end{equation}
Introducing a mass function $M(r)$, the Einstein equation (\ref{eq-1}) reduces to
\begin{equation}
M'(r)=\frac{q^2}{2r^2}-2\mu \frac{q^6}{r^{10}}
\end{equation}
which is easily integrated  to give a newly  metric function
\begin{equation} \label{m-func}
f(r)\equiv 1-\frac{2M(r)}{r}=1-\frac{2m}{r}+\frac{q^2}{r^2}-\frac{4\mu}{9} \frac{q^6}{r^{10}}.
\end{equation}
One recovers the RNBH solution in the limit of $\mu\to 0$. If one includes $\mathcal{F}^2$, instead of $\mathcal{F}^3$, its metric function takes the known form~\cite{Yajima:2000kw}
\begin{equation} \label{g-func}
g(r)=1-\frac{2m}{r}+\frac{q^2}{r^2}-\frac{2\mu}{5} \frac{q^4}{r^{6}}.
\end{equation}
\begin{table}[h]
\begin{tabular}{|c|c|c|c|c|c|c|c|}
  \hline
  $q$& 0.6  & 0.7  & 0.8&0.9&0.95&0.989 \\ \hline
  $\mu$ & (0,$10^{-6}$]&(0,$10^{-5}$]&(0,$10^{-4}$] &(0,$10^{-3}$]&(0,0.003]&(0, 0.01] \\ \hline
   $q$&1&1.001&1.00245&1.004& 1.0062&1.0063 \\ \hline
  $\mu$&(0,0.015] &[0.004,0.016]&[0.01,0.017]&[0.015,0.018]&[0.02,0.02]& N.A. \\ \hline
\end{tabular}
\caption{$\mu$-region for the existence of $r_3(1,q,\mu)$ and $r_4(1,q,\mu)$, depending on the magnetic  charge $q$.  }
\end{table}
Requiring  $f(r)=0$, ten roots are expressed in terms of Mathematica's notation as
\begin{equation}
r_i(m,q,\mu)={\rm Root}(-4\mu q^6+9q^2 \sharp 1^8-18m  \sharp 1^9+9 \sharp 1^{10}\&,i),\quad i=1,2,\cdots,10.
\end{equation}
For $\mu=0.01$ and $m=1$, one finds that  there are four  branches (positive roots) of the  horizon
\begin{equation}
r_{2-}(1,q-,0.01),\quad r_{3}(1,q,0.01),\quad r_{4}(1,q,0.01),\quad r_{2+}(1,q+,0.01)
\end{equation}
and for $\mu=0$, two of $r_6(1,q,0)$ and $r_9(1,q,0)$ describe the outer/inner horizons of the RNBH.
Here, $q-(q+)$ represent $q\in[0,0.989](q\in[0.989,$ ]). Remaining five roots are not  positive functions. We note that  for $\mu=0.1$, there exists  the single branch of $r_2(1,q,0.1)$.
It is important to clarify the $\mu$ and $q$-regions  for the existence of $r_3(1,q,\mu)$ and $r_4(1,q,\mu)$. We compute them numerically because there is no analytic result.
In Table 1, we display the coupling parameter $\mu$-region for the existence of  $r_3(1,q,\mu)$ and $r_4(1,q,\mu)$,  depending on the magnetic charge $q$. For $q=0.6$, its allowed $\mu$-range is very tiny like as $(0,1.73\times 10^{-6}]$.   As $q$ increases from $q=0.6$, the $\mu$-range increases. As $q$ increases from $q=1$, the $\mu$-range becomes smaller and smaller. Its $q$-upper limit is $q=1.0062$ and the corresponding  $\mu$-region is very small like as [0.02093,0.02094] which is less than 0.021 for a given $m=1$. This is why we choose $\mu=0.01$ for realizing four solution  branches of the horizon.  It is worth noting that $\mu$  does not exist beyond  $q=1.0062$. For $q\in[0.989,1.00245]$, $\mu=0.01$ is the upper and lower bounds, guaranteeing the existence of $r_3(1,q,0.01)$ and $r_4(1,q,0.01)$ as is shown in Fig. 1.
\begin{figure*}[t!]
   \centering
   \includegraphics[width=0.5\textwidth]{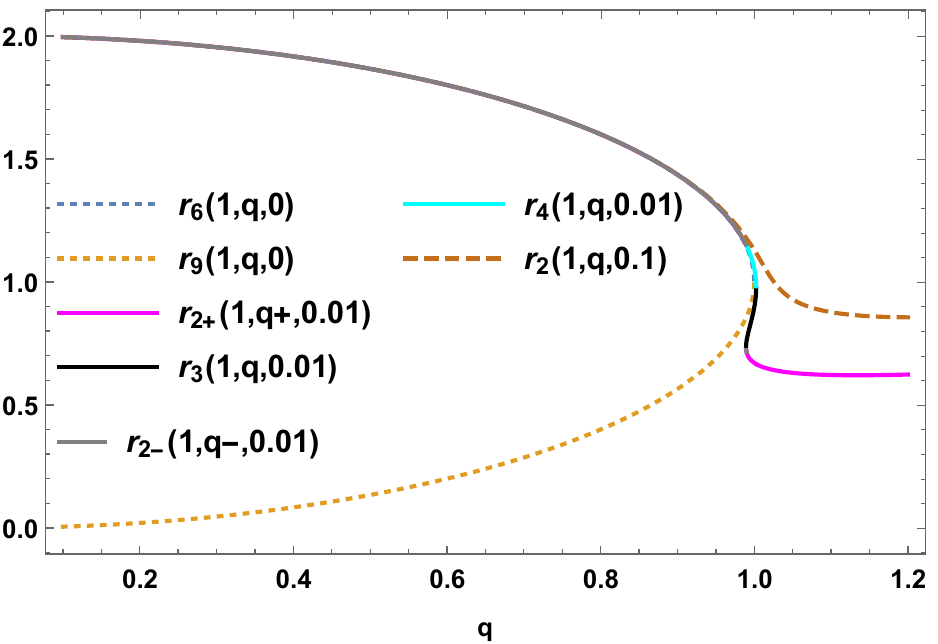}
\caption{ Four branches of the horizon  $r_{2-}(1,q-,\mu=0.01)\ge r_{3}(1,q,0.01)\ge r_{4}(1,q,0.01)\ge r_{2+}(1,q+,0.01)$  as functions of $q$ with $r_{6/9}(1,q,0)$ representing outer/inner horizons for the RNBH. We  note that $r_{2}(1,q,0.1)$ is  defined as  the single horizon without limitation of $q$.   }
\end{figure*}
From  Fig. 1, one finds that $r_{2}(1,q,0.1)$ has a single horizons without limitation of $q$.   An important  point  is
that there is no  theoretical constraint on restricting  the magnetic charge $q$ in the ENED theory.

\section{Photon spheres and shadow radii}

The BH images  indicated that there is a dark central region  surrounded by a bright ring, which are called shadow and photon ring (light ring) of the BH, respectively.
Strong deflection lensing can generate a shadow and relativistic images caused by photons winding several loops around a BH~\cite{Bardeen:1972fi,Bozza:2010xqn}. The photon sphere of a BH  plays an important role in the strong deflection.
We introduce the Lagrangian of the photon to find the photon spheres  around  the NMBH
\begin{equation}
{\cal L}_{\rm LR}=\frac{1}{2}g_{\mu\nu}\dot{x}^\mu\dot{x}^\nu=\frac{1}{2}\Big[-f(r)\dot{t}^2+\frac{\dot{r}^2}{f(r)}+r^2(\dot{\theta}^2+\sin^2\theta \dot{\phi}^2)\Big].
\end{equation}
Taking the light traveling on the equational plane of the  NMBH ($\theta=\pi/2$ and $\dot{\theta}=0$) described by  a spherically symmetric and static metric Eq.(\ref{metric-ansatz}),
two conserved quantities of photon (energy and angular momentum) are given by
\begin{equation}
E=-\frac{\partial {\cal L}_{LR}}{\partial \dot{t}}=f(r)\dot{t},\quad \tilde{\ell}=\frac{\partial {\cal L}_{LR}}{\partial \dot{\phi}}=r^2\dot{\phi}.
\end{equation}
Choosing the null geodesic for the photon ($ds^2=0$) with the affine parameter $\tilde{\lambda}=\lambda \tilde{\ell}$ and impact parameter $\tilde{b}=\tilde{\ell}/E$, its radial equation of motion is  given by
\begin{equation}
\frac{dr}{d\tilde{\lambda}}=\sqrt{\frac{1}{\tilde{b}^2}-\frac{f(r)}{r^2}}.
\end{equation}
Here, the effective potential for a photon takes the form
\begin{equation}
V(r)=\frac{f(r)}{r^2}.
\end{equation}
Requiring  the photon sphere ($\dot{r}=0,~\ddot{r}=0$), one finds two conditions
\begin{equation} \label{cond-LR}
V(r=L)=\frac{1}{2b^2}, \quad V'(r=L)=0,
\end{equation}
where $b$ is  the critical impact parameter and $L$ represents the radius  of unstable photon sphere.
\begin{figure*}[t!]
   \centering
  \includegraphics[width=0.4\textwidth]{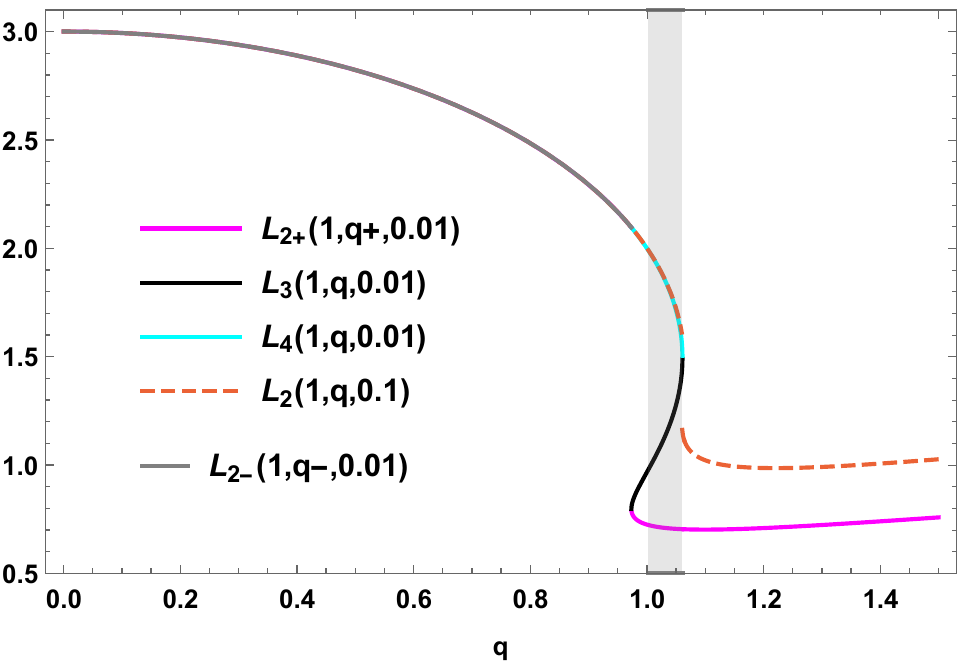}
  \hfill%
    \includegraphics[width=0.4\textwidth]{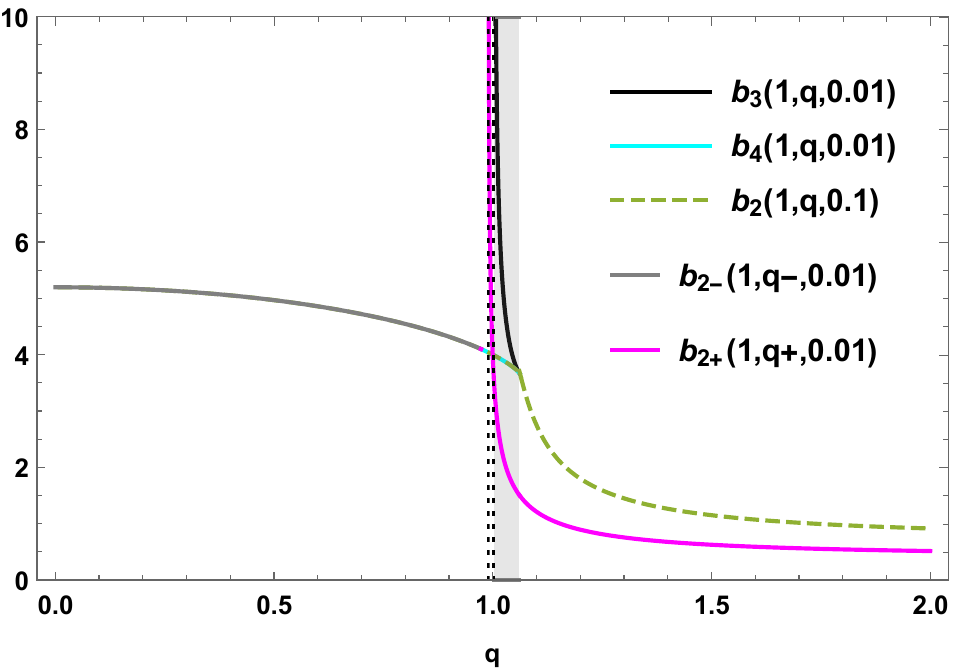}
\caption{ (Left)  Four photon sphere radii  $L_{i}(m=1,q,\mu=0.01)$   for $i=2-,3,4,2+$ and  $L_2(1,q,0.1)$  are  as functions of $q$.  The 3 and 4-branches are  extended to include their NS versions defined in the shaded column ($q\in[1.00245,1.06]$).
(Right) Four critical impact parameters  $b_{i}(m=1,q,\mu=0.01)$   for $i=2_,3,4,2+$ and  $b_2(1,q,0.1)$ is   as functions of $q$.  A shaded column ($q\in[1.00245,1.06]$) includes  their 3, 4-NS branches.  Two dotted lines represent blow-up points of $b_{2+}(1,q+,0.01)$ and $b_3(1,q,0.01)$ at $q=0.989,1.00245$.}
\end{figure*}

 Eq.(\ref{cond-LR}) implies   two relations
 \begin{equation}
 L_i^2=f(L_i)b_i^2,\quad 2f(L_i)-L_if'(L_i)=0.
 \end{equation}
Here, four photon sphere radii and their critical impact parameters ($i=2-,3,4,2+$) are given by
\begin{eqnarray}
&&L_{i}(m,q,\mu), \label{LR} \\
&&b_{i}(m,q,\mu),\label{CI}
\end{eqnarray}
whose explicit forms are too complicated to write down here.

(Left) Fig. 2 shows the photon sphere radii and (Right) Fig. 2 represents shadow radii.
We note that $L_{3}(1,q,0.01)$ and  $L_{4}(1,q,0.01)$ are  present as  connectors appearing  between $L_{2-}(1,q-,0.01)$ and $L_{2+}(1,q+,0.01)$, while  $L_{2}(1,q,0.1)$ is the single function of $q$.
 $L_3(1,q,0.01)$ and $L_4(1,q,0.01)$ are extended to enter their NS branches (3-NS, 4-NS) into this column with $q\in[1.00245,1.006]$.
 Importantly, the dotted lines in (Right) Fig. 2 indicate  the  blow-up  points for $b_{2+}(1,q+,0.01)$ and $b_3(1,q,0.01)$ and a shaded column ($q\in[1.00245,1.06]$) includes  their 3, 4-NS branches.
 However, $L_6(1,q,0)$ and $b_6(1,q,0)$ are not realized into the figure, unlike $L_{RN}(1,q\in[0,1])$ and $b_{RN}(1,q\in[0,1])$ for the RNBH defined as
\begin{equation}
L_{RN}(m,q)=\frac{3m}{2}\Big[1+\sqrt{1-\frac{8q^2}{9m^2}}\Big],\quad b_{RN}(m,q)=\frac{3m(1+\sqrt{1-8q^2/9m^2})}{\sqrt{2+\frac{3m^2(1+\sqrt{1-8q^2/9m^2})}{2q^2}}}.
\end{equation}

\section{Test with EHT results}

Consulting the EHT observation (Keck- and VLTI-based estimates for SgrA$*$~\cite{EventHorizonTelescope:2022wkp,EventHorizonTelescope:2022wok,EventHorizonTelescope:2022xqj}), the  $1\sigma$ constraint on the shadow radius $r_{\rm sh}=b_i$ indicates ~\cite{Vagnozzi:2022moj}
\begin{equation}
4.55\lesssim r_{\rm sh} \lesssim 5.22  \label{KV1}
\end{equation}
and the  $2\sigma$ constraint shows
\begin{equation}
4.21 \lesssim r_{\rm sh} \lesssim 5.56. \label{KV2}
\end{equation}
Let us see (Left) Fig. 3 for an explicit picture to test  with the EHT observation.
For the $2-$-branch and  2-branch with $\mu=0.1$, one has two constraints of the upper limits on its magnetic  charge $q$: $q\lesssim 0.798 (1\sigma)$ and 0.939 $(2\sigma)$, which is the same as $q\lesssim 0.798 (1\sigma)$ and  $0.939(2\sigma)$ for the RNBH with magnetic (electric) charge $q\in[0,1]$.
From (Right) Fig. 3, one observes that two narrow ranges of  $1.022\lesssim q \lesssim 1.03 (1\sigma)$ and  $1.02\lesssim q \lesssim 1.038 (2\sigma)$ exist for the 3-NS branch.
We find that two very narrow ranges of  $0.996\lesssim q \lesssim 0.998 (1\sigma)$ and  $0.995\lesssim q \lesssim 0.999(2\sigma)$ exist for the 2+-branch.
We note that the 4-NS branch is ruled out by the $2\sigma$.
Also, it is worth  noting  that there is no $q>1$ range which constrains its magnetic  charge for $\mu=0.1$  because its critical impact parameter $b_2(1,q,0.1)$ is a monotonically decreasing functions of $q$.
\begin{figure*}[t!]
   \centering
  \includegraphics[width=0.4\textwidth]{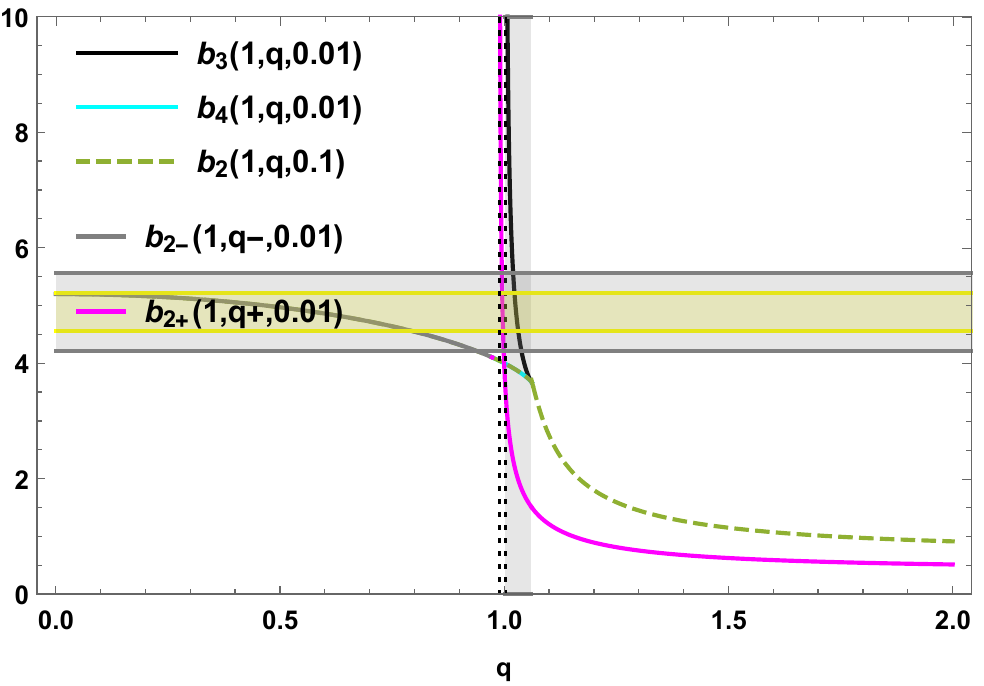}
 \hfill%
    \includegraphics[width=0.4\textwidth]{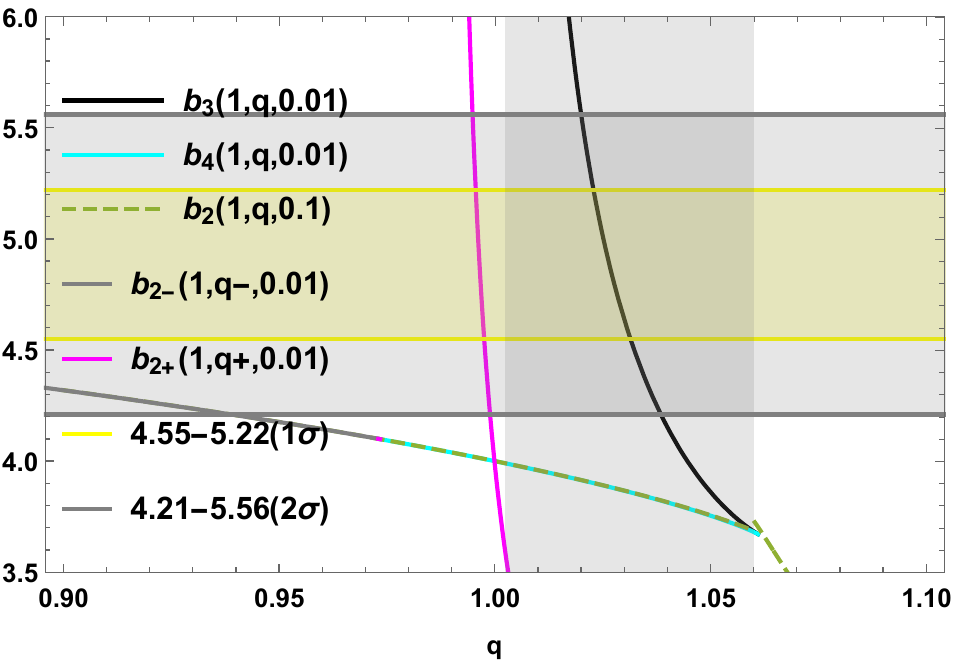}
\caption{(Left) Five  critical impact parameters  $b_i(m=1,q,\mu=0.01)$ for $i=2-,3,4,2+$ and  $b_{2}(1,q,0.1)$  are as functions of $q\in[0,2]$.  There are two  blow-up points (dotted lines) at $q=0.989,1.00245$.
   Here, we introduce $1\sigma$ and $2\sigma$ ranges.
 (Right) Enlarged critical impact parameters $b_i(1,q,0.01)$ for $i=2-,3,4,2+$ and  $b_2(1,q,0.1)$  are  functions of  $q\in[0.9,1.10]$. The shaded column of $q\in[1.00245,1.06]$  includes the 3-NS and   4-NS  branches.   }
\end{figure*}
\section{Geometric  scattering analysis}

Previously, it is known that the critical impact parameters  $b_3(1,q,0.01)$ and $b_{2+}(1,q+,0.01)$ take  peculiar forms, compared with other smooth functions including a continuously decreasing function of $b_{2}(1,q,0.1)$.
These came from the 3-NS and 2+-branches.  To understand them, we need to introduce the scattering picture.
Scattering of a scalar field off a BH is an interesting topic to understand the BH~\cite{Benone:2017hll}.
Scattering and absorption of scalar ﬁelds by various BHs have been  studied for Schwarzschild BH~\cite{Sanchez:1977si}, RNBH~\cite{Crispino:2009ki}, Kerr BH~\cite{Glampedakis:2001cx},  regular BH~\cite{Macedo:2014uga}, and Einstein-Euler-Heisenberg BH~\cite{Olvera:2019unw}.
In the classical (high-frequency) approach, absorption (geometric) cross section of a scalar field  is replaced  by the critical impact parameter as
\begin{equation}
\sigma_{ci}(m,q,\mu)=\pi b^2_i(m,q,\mu),\quad {\rm for}~i=2-,3,4,2+.
\end{equation}
This means that  we may infer their property  by considering the geodesic scattering.  An improvement of the high-frequency cross section was proposed with an inclusion of the  oscillatory part
in the eikonal limit~\cite{Decanini:2011xi}.
There was a recently geometric scattering study  focused on the magnetically charged  BH~\cite{Myung:2025dey}.
\begin{figure*}[t!]
   \centering
  \includegraphics[width=0.4\textwidth]{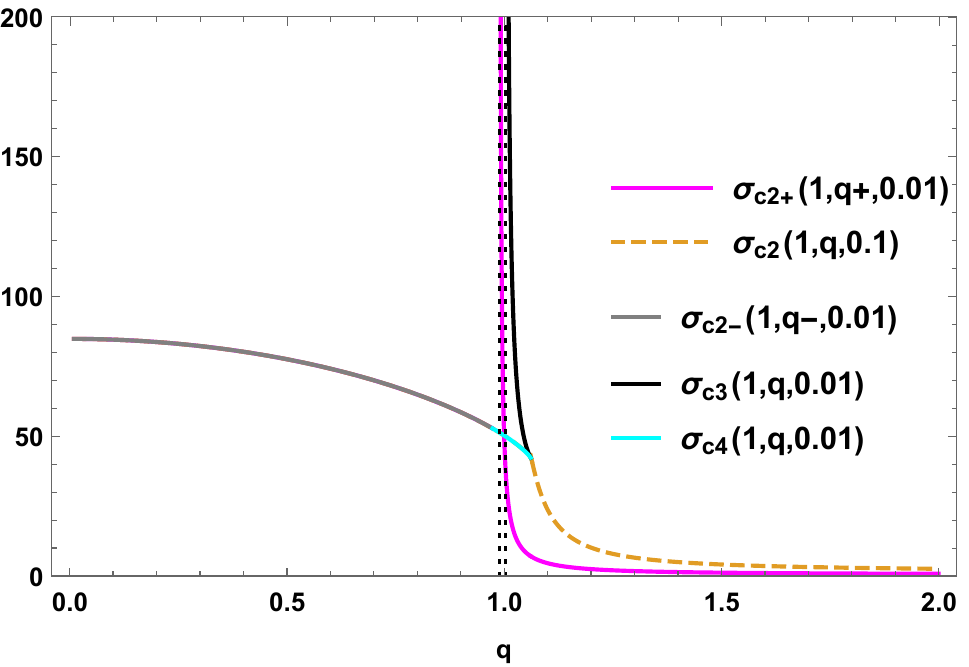}
 \hfill%
    \includegraphics[width=0.4\textwidth]{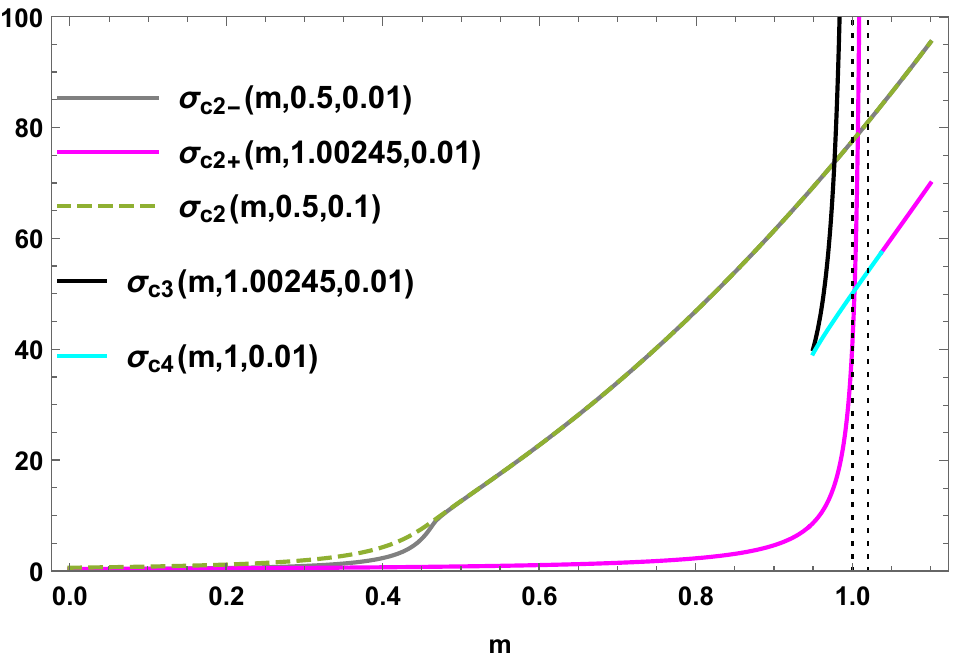}
\caption{(Left).  Five geometric cross sections  $ \sigma_{ci}(m=1,q,0.01)$ for $i=2-,3,4,2+$ and $\sigma_{c2}(1,q,0.1)$ as functions of charge $q\in[0,2]$.
Two dotted lines are  blow-up points at $q=0.989,1.00245$. (Right)   Five geometric cross sections  $ \sigma_{c2-}(m,0.5,0.01), \sigma_{c2}(m,0.5,0.1)$, $ \sigma_{c4}(m,1,0.01)$, $\sigma_{c3}(m,1.00245,0.01)$, and $\sigma_{c2+}(m,1.00245,0.01)$ as functions of mass $m\in[0,1.1]$. The last two are divergent at $m=1.0,1.02$.   }
\end{figure*}

As was shown in Fig. 4, one finds that  $ \sigma_{c2-}(m=1,q-,0.01) \simeq \sigma_{c2}(1,q,0.1)$, implying decreasing functions of  $q$~\cite{Macedo:2014uga}, while   $ \sigma_{c2-}(m,q=0.5,0.01) \simeq \sigma_{c2}(m,0.5,0.1)$, indicating increasing functions of  $m\in [0,1.1]$.  $\sigma_{c4}(1,q,0.01)$ is a decreasing function in $q\in[0.974,1.06]$ whereas  $\sigma_{c4}(m,1,0.01)$ is an increasing function in $m\in[0.943,1.032]$.
This  indicates a promising behavior of  geometric cross sections for the BH with singularity.

However, one finds that  $\sigma_{c2+}(1,q+,0.01)$ and $\sigma_{c3}(1,q,0.01)$  blow up at $q=0.989,1.00245$, and $\sigma_{c3}(m,q=1.00245,0.01)$ and $\sigma_{c2+}(m,q=1.00245,0.01)$ are divergent at $m=1.0, 1.012$.
This may show  an interesting hint of scattering behavior when waves (lights) are scattered off the 3-NS and 2+-branches.  In the case of $b>b_i$ for $i=2-,4$, the particles scatter off the center and the gravitational captures do not happen. For two cases of $q=1.00245$ and $m=1.012$, however,  all particles pass into the 3-NS and 2+ and thus, they all are captured by 3-NS and 2+.
To understand their properties deeply, one needs to compute absorption and scattering of waves with different spins off the 3-NS and 2+.

Finally, we wish to mention two limiting cases of $q$. They include  $\sigma_{c2+}(1,q\to\infty,0.01)=0.54$ and $\sigma_{c2}(1,q\to\infty,0.1)=1.72$,  which means that most particles are scattered off the center (point) and the gravitational captures never happen. The other limit is given by  $\sigma_{c2-}(1,q\to0,0.01)=\sigma_{c2}(1,q\to0,0.1)=84.823$, leading to the geometric cross section for the Schwarzschild BH.

\section{Discussions}

The shadow radii of  various BH and NS found from modified gravity theories  were extensively  used to test the EHT results for SgrA$^*$ BH~\cite{EventHorizonTelescope:2022wkp,EventHorizonTelescope:2022wok,EventHorizonTelescope:2022xqj} and thus, to constrain their hair  parameters~\cite{Vagnozzi:2022moj}.
The magnetically charged BH (\ref{g-func})  was employed to investigating shadow radii~\cite{Allahyari:2019jqz,Vagnozzi:2022moj}.

In this study, we have obtained a newly charged black hole with magnetic charge (NMBH), differing from the known BH solutions.
We tested the NMBH for two coupling constants ($\mu=0.01,0.1$)  with the EHT observation for SgrA$^*$  by computing their shadow radii.
For $\mu=0.01$, we have four branches of $2-,3,4,2+$, while the single branch is allowed for $\mu=0.1$.
In this case, the shadow radius  for the $2-$-branch  is  the  same as that of the RNBH,   while  the 3-NS branch  could be constrained by the EHT observation.
Two narrow allowed ranges of  $1.022\lesssim q \lesssim 1.03 (1\sigma)$ and  $1.02\lesssim q \lesssim 1.038 (2\sigma)$ exist for the 3-NS branch.
Also, two very narrow ranges of  $0.996\lesssim q \lesssim 0.998 (1\sigma)$ and  $0.995\lesssim q \lesssim 0.999(2\sigma)$ are allowed  for the 2+-branch.
For $\mu=0.1$, there exist one single branch  without limitation on magnetic  charge $q$ whose shadow radius is a monotonically decreasing functions of $q$.  This indicates no extremal points. Its shadow radius is the same  as that of the $2-$-branch for $q<1$, but there is no constraints for $q>1$ because $b_2(1,q,0.1)$ is a monotonically decreasing function of $q$.
However, we mention that  even though the 6-branch with $\mu=0$ represents  the outer horizon of the RNBH, its photon sphere and critical impact parameter are not realized into the picture.

From the geometric scattering analysis, we found  that  $ \sigma_{c2-}(m=1,q-,0.01) \simeq \sigma_{c2}(1,q,0.1)$, implying decreasing functions of  $q$~\cite{Macedo:2014uga} while   $ \sigma_{c2-}(m,q=0.5,0.01) \simeq \sigma_{c2}(m,0.5,0.1)$, indicating increasing functions of  $m\in [0,1.1]$.  $\sigma_{c4}(1,q,0.01)$ is a decreasing function in $q\in[0.974,1.06]$, while  $\sigma_{c4}(m,1,0.01)$ is an increasing function in $m\in[0.943,1.032]$.  This  indicates a promising behavior of  geometric cross sections for BH with singularity.
However, we realized that  $\sigma_{c2+}(1,q+,0.01)$ and $\sigma_{c3}(1,q,0.01)$  blow up at $q=0.989,1.00245$,  whereas $\sigma_{c3}(m,q=1.00245,0.01)$ and $\sigma_{c2+}(m,q=1.00245,0.01)$ are divergent at $m=1.0, 1.012$.
This may indicate  an interesting  hint of scattering behavior when waves (lights) are scattered off the 3-NS and 2+-branches.

\vspace{1cm}

{\bf Acknowledgments}
 \vspace{1cm}

 This work was supported by the National Research Foundation of Korea (NRF) grant
 funded by the Korea government (MSIT) (RS-2022-NR069013).

\end{document}